%% file: main.tex
\documentclass[sigconf]{acmart}

\include{commands}

\setcopyright{iw3c2w3}
\begin{document}
\copyrightyear{2021}
\acmYear{2021}
\acmConference[WWW '21]{Proceedings of the Web Conference 2021}{April 19--23, 2021}{Ljubljana, Slovenia}
\acmBooktitle{Proceedings of the Web Conference 2021 (WWW '21), April 19--23, 2021, Ljubljana, Slovenia}
\acmPrice{}
\acmDOI{10.1145/3442381.3449963}
\acmISBN{978-1-4503-8312-7/21/04}

\pagestyle{plain}


\title{Controllable Gradient Item Retrieval}

\author{Haonan Wang$^{12*}$, Chang Zhou$^{2}$, Carl Yang$^{3}$, Hongxia Yang$^{2\dagger}$, Jingrui He$^{1\dagger}$}

\affiliation{
\institution{$^1$ University of Illinois Urbana-Champaign, $^{2}$Alibaba Group, $^{3}$Emory University}
\country{}
}
\affiliation{
\institution{$^1$\{haonan3, jingrui\}@illinois.edu, $^2$\{eirczhou.zc, yang.yhx\}@alibaba-inc.com, $^3$\{j.carlyang\}@emory.edu}
\country{}
}

\setlength{\floatsep}{4pt plus 4pt minus 1pt}
\setlength{\textfloatsep}{4pt plus 2pt minus 2pt}
\setlength{\intextsep}{4pt plus 2pt minus 2pt}
\setlength{\dbltextfloatsep}{3pt plus 2pt minus 1pt}
\setlength{\dblfloatsep}{3pt plus 2pt minus 1pt}
\setlength{\abovedisplayskip}{2pt plus 1pt minus 1pt}
\setlength{\belowdisplayskip}{2pt plus 1pt minus 1pt}

\newcommand{\independent}{\perp\mkern-9.5mu\perp}
\settopmatter{printacmref=false, printfolios=false}

\input{sec-abstract}
\maketitle
{
\renewcommand{\thefootnote}{\fnsymbol{footnote}}
\footnotetext[1]{Work done when he was a research intern at Alibaba Group.}
\footnotetext[2]{Co-corresponding authors.}
}
\input{sec-intro}
\input{sec-model}
\input{sec-exp}
\input{sec-related}
\input{sec-conclusion}
\input{sec-acknowledge}
\bibliographystyle{ACM-Reference-Format}
\bibliography{charleswang}

\end{document}

%% file: commands.tex
\usepackage{booktabs} 
\usepackage{amsmath,url}

\usepackage{amssymb}
 
\usepackage{amsfonts}
\usepackage{multirow}
\usepackage{color}
\usepackage{enumitem}
\usepackage{dsfont}
\usepackage{graphicx}
\usepackage{subcaption}

\newcommand{\mylistbegin}{
  \begin{list}{$\bullet$}
   {
     \setlength{\itemsep}{-2pt}
     \setlength{\leftmargin}{1em}
     \setlength{\labelwidth}{1em}
     \setlength{\labelsep}{0.5em} } }
\newcommand{\mylistend}{
   \end{list}  }

\newcommand{\CGIR}{\textit{CGIR}}

%% file: sec-abstract.tex
\begin{abstract}
In this paper, we identify and study an important problem of gradient item retrieval. We define the problem as retrieving a sequence of items with a gradual change on a certain attribute, given a reference item and a modification text. For example, after a customer saw a white dress, she/he wants to buy a similar one but more floral on it. The extent of "more floral" is subjective, thus prompting one floral dress is hard to satisfy the customer's needs. A better way is to present a sequence of products with increasingly floral attributes based on the white dress, and allow the customer to select the most satisfactory one from the sequence. Existing item retrieval methods mainly focus on whether the target items appear at the top of the retrieved sequence, but ignore the demand for retrieving a sequence of products with gradual change on a certain attribute. To deal with this problem, we propose a weakly-supervised method that can learn a disentangled item representation from user-item interaction data and ground the semantic meaning of attributes to dimensions of the item representation. Our method takes a reference item and a modification as a query. During inference, we start from the reference item and "walk" along the direction of the modification in the item representation space to retrieve a sequence of items in a gradient manner. We demonstrate our proposed method can achieve disentanglement through weak supervision. Besides, we empirically show that an item sequence retrieved by our method is gradually changed on an indicated attribute and, in the item retrieval task, our method outperforms existing approaches on three different datasets.

\end{abstract}
\keywords{information retrieval; recommendation system; weakly-supervised learning; disentangled representation learning; variational autoencoder}

%% file: sec-intro.tex
\section{Introduction}
\label{sec:intro}

\begin{figure}[h!]
\centering
\includegraphics[width=0.48\textwidth]{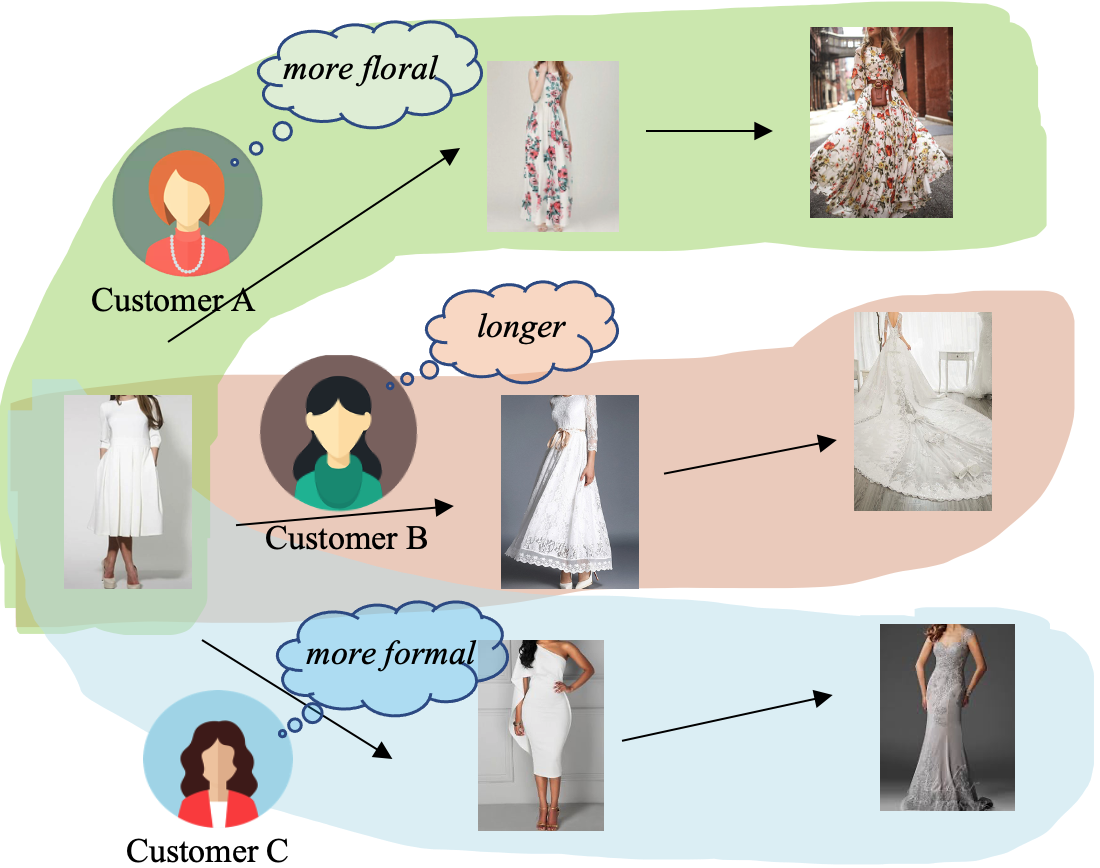}
\caption{\textbf{A motivating example of our proposed framework.} After browsing one white dress, different users want to purchase a dress with some degree of differences in a certain attribute to the white one. }
\label{fig:toy}
\end{figure}

Controllable recommendations are essential for enhancing the customer experience in real-world recommendation scenarios. An example is shown in Figure \ref{fig:toy}: customers are inspired by one white dress and want to purchase a dress with some degree of differences in certain attributes to the white one. In offline shopping, it is easy for the customer to make the salesperson promote a series of products that only differ in certain attributes indicated by the customer in a gradient manner. Then the customer can select the most favorite product from the series of products conveniently. However, it is hard for current recommendation systems to present a sequence of products in a gradient form on a certain attribute based on a reference product. The controllable recommendation as a new type of interaction paradigm can solve the problem. In our work, we define controllable recommendation as a two-stage process. In the first stage, a product will be promoted by the recommendation system along with several modification options for each customer. In the second stage, based on the product and the customer-selected modification, a sequence of products with gradient change on a certain attribute will be retrieved. As this is a new type of interaction with a lot of uncertainty, we need to verify in prototype whether the gradient retrieval is feasible. To make it simple and clean, we keep the discussion of the impact of the customers and the performance of the overall controllable recommendation in the future works. As a first step to approach the controllable recommendation, in this work, we only study the problem of gradient item retrieval with a reference item and a modification as query.

Current methods usually formulate the second stage as a retrieval problem with a text as a query~\cite{Zhen19DSCMR, Vo19TIRG}. Those methods mainly care about whether the target items are retrieved at the top of the retrieved item sequence. Thus, the items in a retrieved item sequence are ranked by the similarities between the input query and items. The demand of retrieving a list of items with gradual change on a certain attribute is largely ignored. The key limitation of these methods is that they only try to model the similarity between the query and target item in their common representation space. In contrast, our method regards a modification text as a "walk" starting from a certain item in the hidden space. By gradually increasing the "step size", a sequence of items can be retrieved in a gradient manner. 

Furthermore, we aim to retrieve a sequence of items with gradual change on a certain attribute with weak supervision.
Specifically, the goal is to retrieve a sequence of items, where the relevance of a certain attribute is in increasing/decreasing order and other attributes keep the same level. Note the desired attributes (e.g. "floral", "formal") and modification actions ("more" or "less") are indicated by a modification text. To solve the problem, we propose a novel \textbf{C}ontrollable \textbf{G}radient \textbf{I}tem \textbf{R}etrieval framework, called \CGIR~, which learns disentangled item representations with semantic meanings. In the training stage, we only need to know whether a certain product has this attribute or not in order to ground the semantic meanings of each attribute to dimensions of the factorized representation space. This type of weak supervision alleviates the burden of obtaining hand-labeled item sequences with gradual change for an attribute. Thanks to the disentanglement property of learned item representations, we can modify the value on dimensions associated with an indicated attribute to form queries without affecting irrelevant attributes. In the inference stage, by using the queries with different modification strength, a sequence of items can be retrieved in a gradient manner.

Unlike previous unsupervised disentanglement methods which have been demonstrated to rely heavily on model inductive bias and require careful supervision-based hyper-parameter tuning~\cite{Locatello19challengedisentangle}, in this work, we propose a weakly supervised setting to learn disentangled item representations. Specifically, to achieve disentanglement, our method grounds the semantic meanings of attributes to different dimensions of the factorized representation. Following the previous discussion about disentanglement~\cite{shu20disentangleguarantee}, we decompose disentanglement into two distinct concepts: \textit{consistency} and \textit{restrictiveness}. Specifically, \textit{consistency} means only when the hidden factor of one attribute changes, the attribute will change accordingly; and \textit{restrictiveness} means when one hidden factor changes, irrelevant attributes will keep the same~\cite{shu20disentangleguarantee}. By enforcing the disentangled factors to match the oracle hidden factors and encoding them into separate dimensions of representation, our proposed method can satisfy the two properties, which allow us to retrieve items with gradual changes along a certain attribute by tuning the value of relevant dimensions. 

To summarize, the main contributions of this paper are:
\begin{itemize}[leftmargin=15pt]
    \item We identify and define the task of gradient item retrieval.
    \item For the first time, we propose a weakly-supervised disentanglement framework that can ground semantic meanings to dimensions of a disentangled representation space.
    \item We demonstrate that our weakly-supervised method can achieve the desired representation disentanglement with semantic meanings, and
    empirically show that our method can achieve gradient retrieval on both public and industrial datasets.
\end{itemize}

The rest of this paper is organized as follows. The proposed \CGIR~ is introduced in Section 2. Qualitative and quantitative experiments are given in Section 3. Section 4 reviews the related work.  Finally, we conclude this work in Section 5.

%% file: sec-model.tex
\section{Proposed CGIR Method}
\label{sec:method}
In this section, we first formally define the notation and the gradient item retrieval problem. Then we introduce the proposed framework, followed by discussions about how the proposed method can learn the disentangled item representations with semantic meanings. After that, we show that our weakly-supervised method can achieve disentangled representation with consistency and restrictiveness theoretically.

\begin{figure*}[h!]
\centering
\includegraphics[width=0.9\textwidth]{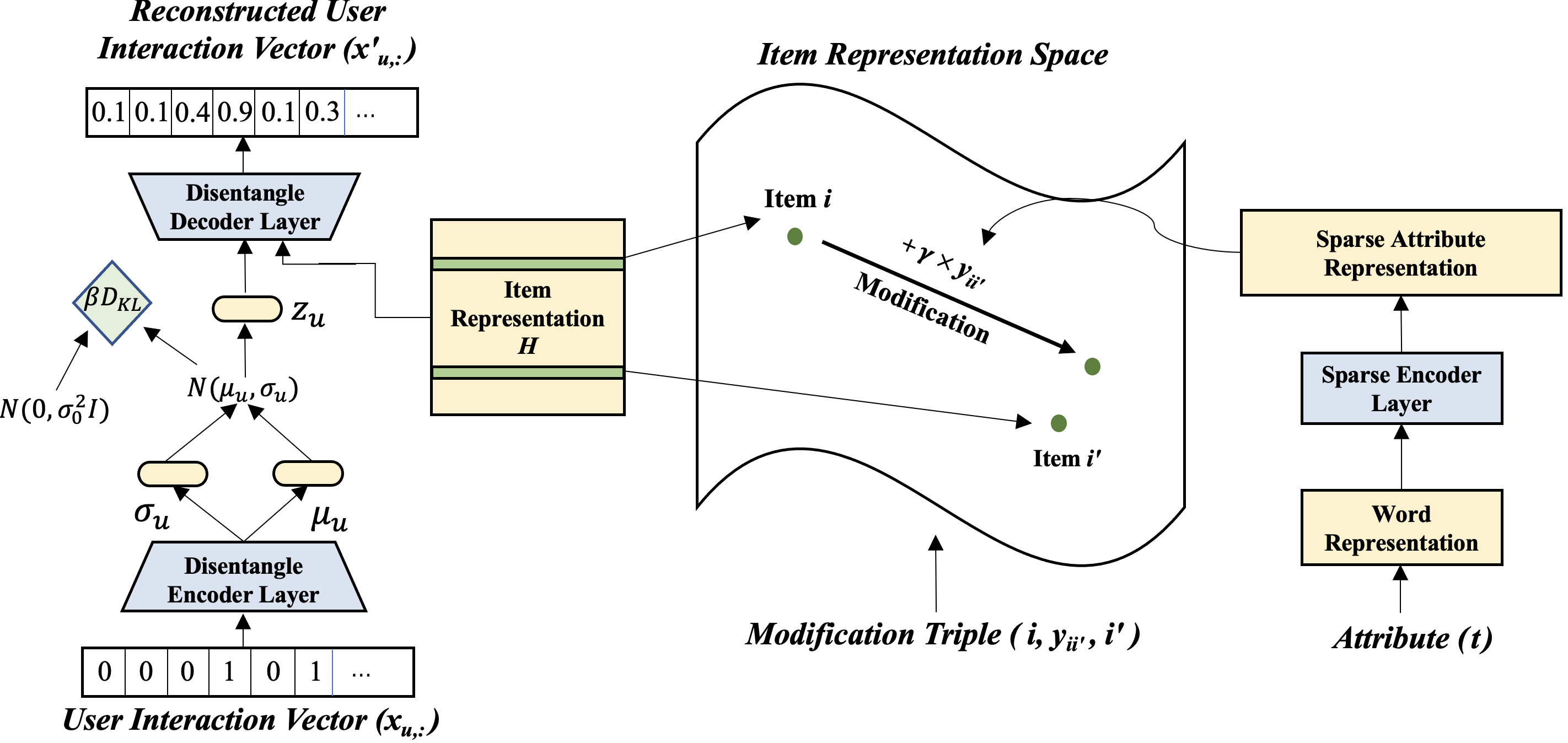}
\caption{Overview of our proposed \CGIR~ framework. It includes three major parts -- the left is for disentangled item representation, the right part aims at enforcing representation of attributes to be sparse, and the middle part is for aligning the disentangled item representation space and the sparse word representation space. They are trained in an end-to-end manner.}
\label{fig:model}
\end{figure*}

\subsection{Notation and Problem Formulation}
\textbf{Notation} In this problem, we are provided with a set of users $\mathcal{U}$, a set of items $\mathcal{I}$, a set of attribute strings $\mathcal{T}$, interaction data $\mathbf{X}$ between users and items, and item-attribute relation data $\mathbf{A}$ between attributes and items. Specifically, the interaction data $\mathbf{X}$ consists of the interactions between $N$ users and $M$ items. An interaction between user $u$ and item $i$ is denoted by $x_{u,i} \in \{0,1\}$, where $x_{u,i}=1$ indicates that user $u$ adopts item $i$, whereas $x_{u,i}=0$ means there is no recorded interaction between them. For convenience, we use $\mathbf{x}_{u,:}$ to represent the items adopted by user $u$ and $\mathbf{x}_{:,i}$ to denote the users who interacted with item $i$. The item-attribute relation data $\mathbf{A}$ consists of relations between $M$ items and $T$ attributes, $T = |\mathcal{T}|$. If item $i$ has attribute $t$, then $a_{i,t} = 1$, otherwise $a_{i,t} = 0$. The attribute vector of item $i$ is denoted as $\mathbf{a}_{i,:}$. Besides, the attribute difference data $\mathbf{Y}$ is composed of attribute difference vector $\mathbf{y}_{i,i'} = \mathbf{a}_{i,:} - \mathbf{a}_{i',:}$ , $\mathbf{y}_{i,i'} \in R^{T}$. Each element of the difference vector $y_{i,i'}^t \in \{-1,0,1\}$ indicates the difference between item $i$ and $i'$ on a certain attribute $t$. Triple data $\mathcal{D}$ is constructed using previously mentioned data and it is composed of $(i, \mathbf{y}_{i,i'}, i')$ triples where $i$ denotes reference item, $\mathbf{y}_{i,i'}$ denotes modification and $i'$ is the desired target item.

\textbf{Problem Definition} We define the gradient item retrieval problem as follows: \textit{based on a reference item and a modification, retrieve a sequence of items in which relevance for a certain desired attribute is in increasing or decreasing order, and relevance for other attributes remains the same}. To make it simple, we consider that a query consists of a reference item and a modification about only one attribute. Note that, if multiple attributes are required to be modified, we can apply the atomic modification several times. Mathematically, we define the query as $(i, \alpha t)$ where $i$ indicates the reference item, $\alpha \in \{1,-1\}$ is the modification action and $t$ is the desired modification attribute. Note that there is a bijection between $\alpha$ and the modification words "more" and "less". For the gradient item retrieval problem, it can be defined as: for a query $(i, \alpha t)$ and its corresponding retrieval sequence $Seq\text{-}i$, we want to maximize the probability of the sequence satisfying the constraint: $\alpha \cdot relevance(Seq\text{-}i@k,t) < \alpha \cdot relevance(Seq\text{-}i@k+1,t)$ and $ relevance(Seq\text{-}i@k,t') = relevance(Seq\text{-}i@k+1,t')$, for any other $ t' \in \mathcal{T}, t' \neq t$, where the $relevance$ function measures the relevance score between a retrieved item $Seq\text{-}i@k$ and a certain attribute.

\subsection{Proposed Framework of CGIR}
The general framework of the proposed method is shown in Figure \ref{fig:model}. It includes three major parts. 

The left part is designed based on Variational Autoencoder framework~\cite{KingmaW13VAE}, which learns a disentangled item representation from user activities. For each user $u$, we encode the interaction vector $x_{u,:}$ to the user hidden representation $\mathbf{z}_{u} \in R^{D}$. After calculating the interaction probability between user $u$ and all items $\mathbf{H} \in R^{M \times D}$, we reconstruct the interaction vector $x'_{u,:} \in R^{M}$. The reconstruction loss can be calculated between user interaction vector $x_{u,:}$ and its relevant reconstructed vector $x'_{u,:}$. The disentanglement loss is computed using the mean $\mu_{u}$ and variance $\sigma_{u}$.
We keep the dimensionality of $\mu_{u}$ and  $\sigma_{u}$ the same as $\mathbf{z}_{u}$.

The right part aims to encode attribute strings to a space where attribute representations are sparse. In that space, each representation of an attribute string has only a few activated dimensions. Our intuition is, for each item, the information of its disentangled representation includes the information of all its attributes. Therefore, each attribute representation should only correspond to some dimensions of the disentangled representation. The input of the sparse encoder model is pre-trained word vectors. We use GloVe~\cite{Pennington14glove} as initial features for English words and pre-trained Chinese Word Vectors~\cite{Shen18weibo} as initial features for Chinese characters~(attribute data of Alishop-attribute dataset is in Chinese). If an attribute only has one word or phrase, we use the relevant sparse word representation as the attribute representation. For attributes including multiple words, a sum pooling is applied over sparse representations of words to obtain the attribute representation.

The middle part is for aligning the disentangled item representation space and the sparse attribute representation space. By leveraging the VAE framework, representations are factorized, where dimensions tend to be independent~\cite{Locatello18ChallengeDisentangle}. However, the meaning of each dimension or the composition of some dimensions remains unclear. The goal of this part is to ground the semantic meanings of attributes to dimensions of factorized item representations. 

To achieve the goal, one direct way is to leverage the item-attribute relation data $\mathbf{A}$, which adopted by some existing GAN-based methods \cite{CycleGAN2017, lee2018diverse, shen2020interpreting, jahanian2020steerability, zhou2020ganbased}. However, those methods ignore the relationship between items, which contradicted with the essence of item retrieval. Instead, we implicitly align the two space by minimizing the distance between the target item $i'$ and the modification result which computed by adding a correct modification $\mathbf{y}_{i,i'}$ on the reference item $i$. Note, overlapping is allowed between corresponding dimension sets of two attributes, because two attributes may have the same semantic primitives which are separately encoded into different dimensions of item representation. And to keep a linear relationship in the hidden space, we directly add an item representation and an attribute representation without any non-linear transformation. The coefficient $\gamma$ controls the strength of modification. In the training stage, we set $\gamma = 1$ since we only use the information of whether one item has a certain attribute or not. During the inference stage, in order to retrieve a sequence of items in a gradient manner, we change the strength coefficient $\gamma$ by increasing a fraction number at each step and keep the top one retrieved item for each step to form the retrieval sequence. The three parts are trained in an end-to-end manner.

\subsection{Weakly-Supervised Disentangled Representation Learning with Semantic Meaning}
\begin{figure}[h]
\centering
\includegraphics[width=0.35\textwidth]{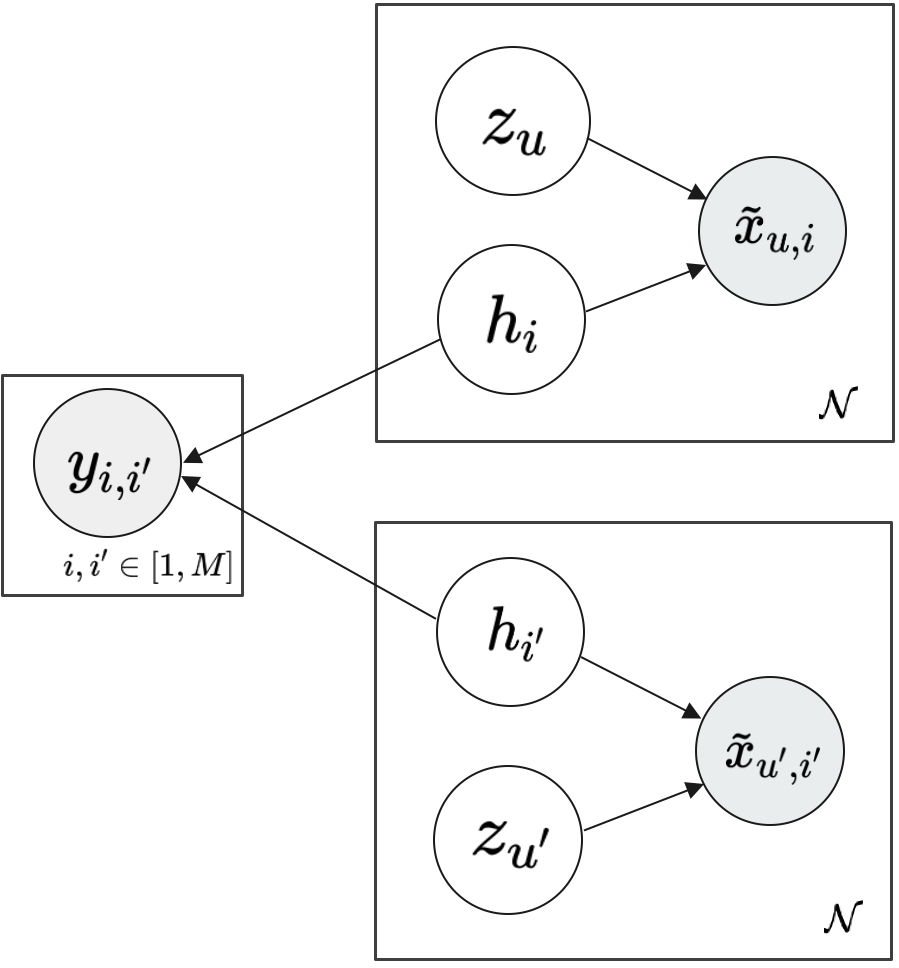}
\caption{\textbf{The decoder model,} $p(\mathbf{X,Y}|\mathbf{Z,H})$.}
\label{fig:pgm1}
 \end{figure}

\textbf{Weakly-Supervised Variational Auto-Encoder.} 
we leverage the VAE framework \cite{KingmaW13VAE} to  enforce item representations to be factorized. And, to involve the information of attribute data, as we stated in the previous section, we model the relation between item pairs and attributes instead of item-attribute data. Specifically, we model the joint distribution of observed variables $\mathbf{X}$ and $\mathbf{Y}$ by joint distribution $p_{\theta}(\mathbf{X}, \mathbf{Y})$ where $\theta$ denotes parameters of \CGIR~. Our generative model assumes that the observed data are generated from the following distribution:
\begin{equation}
\begin{split}
p_{\theta}(\Tilde{\mathbf{X}}, \Tilde{\mathbf{Y}}) &= \iint p_{\theta}(\mathbf{\Tilde{X},\Tilde{Y}}|\mathbf{Z,H}) p_{\theta}(\mathbf{Z,H})  \,d\mathbf{Z} \,d\mathbf{H}
\end{split}
\label{equ:joint}
\end{equation}
$\mathbf{\Tilde{X}}$ and $\mathbf{\Tilde{Y}}$ are variables sampled from a distribution parameterized by hidden variables $\mathbf{Z} \in R^{N \times D}$ and $\mathbf{H} \in R^{M \times D}$. The meanings of $\mathbf{Z}$ and  $\mathbf{H}$ are described in the previous subsection. As shown in Figure \ref{fig:pgm1}, $\mathbf{\Tilde{X},\Tilde{Y}}$ are independent when conditional on $\mathbf{Z}$ and $\mathbf{H}$. Therefore, we have, 
\begin{equation}
\begin{split}
p_{\theta}(\Tilde{\mathbf{X}}, \Tilde{\mathbf{Y}}) = \iint p_{\theta}(\mathbf{\Tilde{X}}|\mathbf{Z,H}) p_{\theta}(\mathbf{\Tilde{Y}}|\mathbf{Z,H})  \,d\mathbf{Z} \,d\mathbf{H}
\end{split}
\label{equ:joint_split}
\end{equation}
We assume interactions between users and items are independent and identically distributed ($i.i.d.$), and vectors in attribute difference data are also $i.i.d.$. Therefore, for the two terms in equation \ref{equ:joint_split}, we have $p_{\theta}(\mathbf{\Tilde{X}}|\mathbf{Z,H}) = \prod_{u,i}p_{\theta}(\Tilde{x}_{u,i}|\mathbf{z}_u, \mathbf{h}_i)$ and  $p_{\theta}(\Tilde{\mathbf{Y}}|\mathbf{Z,H}) = \prod_{i,i'}p_{\theta}(\mathbf{\Tilde{y}}_{i,i'}|\mathbf{h}_i, \mathbf{h}_{i'})$ separately. Following the paradigm of variational autoencoder (VAE) \cite{ChenB20PairwiseVAE, mazhou0Y019MacridVAE}, we introduce a variational dsitribution to alleviate computational burden of integral of equation\ref{equ:joint_split} and maximize the lower bound of $\ln p_{\theta}(\Tilde{x}_{u,i}, \mathbf{\Tilde{y}}_{i,i'}) $ by:

\begin{equation}
\begin{aligned}
\ln p_{\theta}(\Tilde{x}_{u,i}, & \mathbf{\Tilde{y}}_{i,i'}) \geq \mathbb{E}_{q_{\theta}{(\mathbf{z}_{u}, \mathbf{h}_{i}|{x}_{u,i}})} \big[\ln p_{\theta}(\Tilde{x}_{u,i} |\mathbf{z}_{u}, \mathbf{h}_{i})\big] \\
& - \mathcal{D}_{KL}\big(q_{\theta}(\mathbf{z}_{u}, \mathbf{h}_{i}|{x}_{u,i}) || p(\mathbf{z}_{u}, \mathbf{h}_{i}) \big) \\
&+ \mathbb{E}_{q_{\theta}{(\mathbf{z}_{u}, \mathbf{h}_{i}|{x}_{u,i}}), q_{\theta}{(\mathbf{z}_{u}, \mathbf{h}_{i'}|{x}_{u,i'}})} \big[\ln p_{\theta}(\mathbf{\Tilde{y}}_{i,i'} |\mathbf{h}_{i}, \mathbf{h}_{i'})\big]. 
\end{aligned}
\label{equ:vae_lb}
\end{equation}

\begin{figure}[h]
\centering
\includegraphics[width=0.20\textwidth]{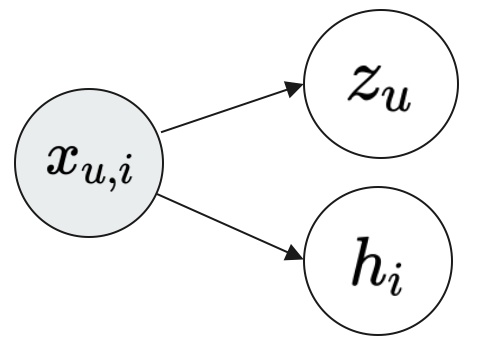}
\caption{\textbf{The encoder model,} $p(\mathbf{Z,H}|\mathbf{X})$.}
\label{fig:pgm2}
\end{figure}

The expectation $\mathbb{E}_{q_{\theta}{(\mathbf{z}_{u}, \mathbf{h}_{i}|x_{u,i}})}[\cdot]$ is still intractable. As shown in figure \ref{fig:pgm2}, we have $\mathbf{z}_{u} \perp \mathbf{h}_{i} | x_{u,i}$, according to the Common cause decomposition of graphical models\cite{Buntine11PGM}. Therefore, we have the following decomposition:

\begin{equation}
\begin{aligned}
q_{\theta}(\mathbf{z}_u,\mathbf{h}_i,|x_{u,i}) = q_{\theta}(\mathbf{z}_u,|x_{u,i}) q_{\theta}(\mathbf{h}_i,|x_{u,i}). \end{aligned}
\label{equ:demopose1}
\end{equation}

Instead of computing $\mathbb{E}_{q_{\theta}{(\mathbf{z}_{u}, \mathbf{h}_{i}|x_{u,i}})}[\cdot]$ directly, we use the Gaussian re-parameterization trick\cite{KingmaW13VAE} to solve $\mathbb{E}_{q_{\theta}(\mathbf{z}_u,|x_{u,i}) q_{\theta}(\mathbf{h}_i,|x_{u,i})}[\cdot]$.

\textbf{Factorization via Regularization.} A natural strategy to encourage factorization is to force statistical independence between dimensions. 
As demonstrate in the previous work \cite{Higgins17betaVAE}, if the prior satisfies factorization, penalizing the Kullback-Leibler term of equation \ref{equ:vae_lb} would encourage independence between the dimensions. In here, we choose two standard multivariate normal distributions as priors for $\mathbf{z}_u$ and $\mathbf{h}_i$. For the Kullback-Leibler divergence part of equation \ref{equ:vae_lb},  we can decompose it as:

\begin{equation}
\begin{split}
&\mathcal{D}_{KL}\big(q_{\theta}(\mathbf{z}_{u}, \mathbf{h}_{i}|x_{u,i}) || p(\mathbf{z}_{u}, \mathbf{h}_{i}) \big) \\
&= \mathcal{D}_{KL}\big(q_{\theta}(\mathbf{z}_{u}|x_{u,i}) q_{\theta}(\mathbf{h}_{i}|x_{u,i})|| p(\mathbf{z}_{u}) p ({ \mathbf{h}_{i}})\big)\\
&= \mathcal{D}_{KL}\big(q_{\theta}(\mathbf{z}_{u}|x_{u,i})||p(\mathbf{z}_{u})\big) + \mathcal{D}_{KL}\big(q_{\theta}(\mathbf{h}_{i}|x_{u,i})||p(\mathbf{h}_{i})\big) 
\end{split}
\label{equ:kl_decompose1}
\end{equation}

The two KL terms in equation \ref{equ:kl_decompose1} aim at enforcing factorization of user and item representations separately.
Due to the time-efficient requirement of recommendation system, we keep a representation table for items, instead of inferring them from interaction matrix at each time. Therefore, we only keep the first term of equation \ref{equ:kl_decompose1} in the final objective. Although this simplification has been used in the previous work\cite{mazhou0Y019MacridVAE}, we also empirically show  that this simplification can enforce item representations to be factorized in our experiments. Besides, We follow $\beta$-VAE\cite{Higgins17betaVAE} to strengthen the KL divergence by a factor of $\beta$.

\textbf{Geometric Relationship of Item Representation.} As shown in the middle part of Figure \ref{fig:model}, to implicitly align item space and attribute space, we leverage the geometric relationship between items. For a reference-target item pair, their distance will be minimized when a correct modification is added on the reference item. Based on the intuition, we define the third term of equation \ref{equ:vae_lb} as:

\begin{equation}
\begin{aligned}
&p_{\theta}(\mathbf{\Tilde{y}}_{i,i'}|\mathbf{h}_{i}, \mathbf{h}_{i'}) =\\
&\frac{q_{\theta}(\mathbf{h}_{i'}|\mathbf{x}_{:,i'}) \big(q_{\theta}(\mathbf{h}_{i}|\mathbf{x}_{:,i}) + \gamma \cdot \sum_{t \in \mathcal{T}} \Tilde{y}_{i,i'}^{t} \cdot F_{\theta}(t)\big)}{\sum_{j' \in [1,M]} q_{\theta}(\mathbf{h}_{i'}|\mathbf{x}_{:,i'})   \big(q_{\theta}(\mathbf{h}_{i}|\mathbf{x}_{:,i}) + \gamma \cdot \sum_{t \in \mathcal{T}} \Tilde{y}_{i,j'}^{t} \cdot F_{\theta}(t)\big)}
\label{equ:contrastive}
\end{aligned}
\end{equation}

In whole, $\gamma \cdot \sum_{t \in \mathcal{T}} \Tilde{y}_{i,i'}^{t} \cdot F_{\theta}(t)$ represents the modification $\mathbf{y}_{i,i'}$ scaled by a factor $\gamma$. During training stage, we set the modification strengthen coefficient $\gamma$ equals one. And during inference, $\gamma$ will be gradually changed to retrieve item in gradient manner. The $\Tilde{y}_{i,i'}^{t}$ indicates the modification direction for attribute $t$, $F_{\theta}(\cdot):R^{K} \rightarrow R^{D}$ is the sparse attribute encoder which encode the attribute $t$ to a sparse representation. The equation \ref{equ:contrastive} represents the probability of one triple $(i, \mathbf{y}_{i,i'}, i')$ in $\mathcal{D}$. To align two representation spaces, we maximize the equation \ref{equ:contrastive}.

\textbf{Sparse Attribute Representation} Following our intuition that one item's attribute has less information than the whole item and should only be grounded to part of disentangled item representations, we enforce the attribute representation to be sparse before the alignment of attribute and item representation. Function $F_{\theta}(\cdot)$ is an attribute encoder which maps a attribute string to a sparse representation space. Specifically,

\begin{equation}
\begin{aligned}
F_{\theta}(t) = \sum_{w \in \mathcal{W}(t)} f_{\theta}(w)
\end{aligned}
\end{equation}

where $\mathcal{W}(t)$ represents the set of words used in attribute string $t$. Function $f_{\theta}(\cdot)$ upscale the word representation to another representation space. To enforce the word representation has only a few activated dimensions a sparse loss ($\textit{SL}$) is applied:

\begin{equation}
\begin{aligned}
\textit{SL} = \frac{1}{D} \sum_{d=1}^D &\Big( \text{max}  (\frac{1}{|\mathcal{W}|} \sum_{w \in \mathcal{W}} f^{d}_{\theta}(w) - \rho, 0) ^2 \\
&+ \frac{1}{|\mathcal{T}|} \sum_{w \in \mathcal{W}}  f^{d}_{\theta}(w) \times (1- f^{d}_{\theta}(w))  \Big).
\label{equ:sparseloss}
\end{aligned}
\end{equation}

The first term is an Average Sparsity Loss (ASL) which penalizes any deviation of the observed average activation value $f^{d}_{\theta}(w)$ from the desired average activation value $\rho$ which is usually set to a small value. The second term is a Partial Sparsity Loss (PSL) that facilitates the value of each dimension of $f_{\theta}(w)$ to be close to either 0 or 1\cite{Subramanian18SPINE}.

\textbf{Overall Objective Function} The above equations bring us to the following training objective. Parameter $\theta$ is optimized by maximizing the objective:

\begin{equation}
\begin{split}
\mathbb{E}_{ p_{data}(\mathbf{X})} & \Big[ \mathbb{E}_{q_{\theta}{(\mathbf{z}_{u}, \mathbf{h}_{i}|x_{u,i}})} \big[\ln p_{\theta}(\Tilde{x}_{u,i} |\mathbf{z}_{u}, \mathbf{h}_{i})\big] \\
& - \mathcal{D}_{KL}\big(q_{\theta}(\mathbf{z}_{u}|{x}_{u,i})||p(\mathbf{z}_{u})\big) \\
&+ \mathbb{E}_{q_{\theta}{(\mathbf{z}_{u}, \mathbf{h}_{i}|{x}_{u,i}}), q_{\theta}{(\mathbf{z}_{u}, \mathbf{h}_{i'}|{x}_{u,i'}})} \big[\ln p_{\theta}(\mathbf{\Tilde{y}}_{i,i'} |\mathbf{h}_{i}, \mathbf{h}_{i'})\big] \Big] \\
&-\frac{1}{D} \sum_{d=1}^D \Big( \text{max}  \big(\frac{1}{|\mathcal{W}|} \textstyle\sum_{w \in \mathcal{W}} f^{d}_{\theta}(w) - \rho, 0\big)^2 \\
&+ \frac{1}{|\mathcal{T}|} \textstyle\sum_{w \in \mathcal{W}}  f^{d}_{\theta}(w) \times \big(1- f^{d}_{\theta}(w) \big)  \Big).
\end{split}
\label{equ:whole_obj}
\end{equation}

\subsection{Disentanglement with Guarantee}
As demonstrated by Locatello et al.\cite{Locatello19challengedisentangle}, VAE-based unsupervised learning methods fundamentally cannot achieve disentanglement without model inductive biases. Therefore, a natural question is can our method deliver a disentanglement without the help of model inductive bias? Shu et al.\cite{shu20disentangleguarantee} gives a theoretical analysis which shows disentanglement can be achieved with guarantee under proper weak supervision. Within their analysis framework, three types of weakly supervised settings were considered, which are restricted labeling, matching pairing, and rank pairing. In our case, attributes of items are considered as hidden factors. For item $i$ and item $i'$, we construct the attribute difference vector $\mathbf{y}_{i,i'}$ by comparing them under each attribute $t$, $y_{i,i'}^t = a_{i,t} - a_{i',t}$. If attribute $t$ belongs to item $i$ but not for item $i'$, then the ranking of $i$ is higher than $i'$ and $y_{i,i'}^t$ equals $1$. Therefore, the $(i,\mathbf{y}_{i,i'},i')$ triple data can be understood as a special type of ranking-pair where the ranking is binarized and semantic meaningful. Then according to the \textit{Weak Supervision Disentanglement Theorem}\cite{shu20disentangleguarantee}, the disentangled representation learned under three types of weak supervision is distribution-matching an oracle disentangled representation in which the consistency property of hidden factors, considered by weak supervision signal, can be guaranteed. In our setting, we consider the ranking of one attribute between two items at each triple, because of the restriction $\sum_{t\in\mathcal{T}} |{y}^t_{i,i'}| = 1$. Further, empirically we have $\sum_{i,i' \in [1,N]} |{y}^t_{i,i'}| > 1, \forall t \in \mathcal{T}$ which means all attributes are considered by the weak supervision signal. Further, the consistency of all attributes can be guaranteed. By the Full Disentanglement Rule\cite{shu20disentangleguarantee}, the consistency of all factors further implies the restrictiveness property is guaranteed in disentangled representation.

\begin{equation}
\begin{split}
\bigwedge_{t \in \mathcal{T}} C(t) \iff  \bigwedge_{t \in  \mathcal{T}} D(t), \bigwedge_{t \in \mathcal{T}} D(t) \iff  \bigwedge_{t \in  \mathcal{T}} R(t)
\end{split}
\end{equation}

where $C(t)$ denotes the consistency of hidden factor $t$, $R(t)$ denotes restrictiveness of hidden factor $t$ and $D(t)$ denotes the disentanglement of hidden factor $t$. 

%% file: sec-exp.tex
\section{Experiments}

\begin{table*}[ht]
\centering
\small
\setlength{\tabcolsep}{4pt}{
\begin{tabular}{l | llll | llll | llll }
  \bottomrule
& \multicolumn{4}{c|}{\bf ALiShop-tag} & \multicolumn{4}{c|}{\bf ML-25M}& \multicolumn{4}{c}{\bf ML-20M}\\
\hline
Models &{\bf Hit@20}&{\bf Hit@50} &{\bf MRR}&{\bf MGS} &{\bf Hit@20}&{\bf Hit@50} &{\bf MRR}&{\bf MGS} &{\bf Hit@20}&{\bf Hit@50} &{\bf MRR}&{\bf MGS} \\
\hline
CBIR  & 0.0211 & 0.0409 & 0.01601 & 0.1811 & 0.2651 & 0.3328 & 0.2116 & 0.1629 & 0.3127 & 0.4638 & 0.2481 & 0.1494 \\
DSCMR & 0.0239 & 0.0591 & 0.01769 & 0.1934 & 0.2974 & 0.3471 & 0.2292 & 0.1683 & 0.3271 & 0.4810 & 0.2622 & 0.1543 \\
TIRG  & 0.0581 & 0.0831 & 0.02418 & 0.2364 & 0.4328 & 0.4801 & 0.3094 & 0.1903 & 0.4733 & 0.5497 & 0.3286 & 0.2085 \\
\hline
\CGIR & \textbf{0.0626} & \textbf{0.1019} & \textbf{0.02638} & \textbf{0.2796} & \textbf{0.4412} & \textbf{0.4891} & \textbf{0.3164} & \textbf{0.2588} & \textbf{0.4986} & \textbf{0.5572} & \textbf{0.3374} & \textbf{0.2359} \\
\CGIR~ w/o VAE & 0.0572 & 0.0810 & 0.02421 & 0.2314 & 0.4286 & 0.4731 & 0.2981 & 0.2094 & 0.4729 & 0.5334 & 0.3196  & 0.1938 \\
\CGIR~ w/o Sparse &  0.0628 &  0.1021 &  0.02641 & 0.2607 &  0.4462 &  0.4905 &  0.3188 & 0.2361 &  0.4990 &  0.5578 &  0.3375 & 0.2162\\ 
\toprule
\end{tabular}}
\vspace{-3pt}
\caption{Gradient Item Retrieval Performance Evaluation on three datasets. } 
\vspace{-13pt}
\label{tab:IR_performance}
\end{table*}

We evaluate \CGIR~ on real-world datasets with the aim to answer the following research questions (RQs): \\
\textbf{RQ 1} Does \CGIR~ achieve gradient item retrieval? \\
\textbf{RQ 2} Does \CGIR~ outperform other competitors in the item retrieval task?\\
\textbf{RQ 3} Can \CGIR~ achieve factorized item representation?\\


\begin{table}[H]
\centering
\setlength{\tabcolsep}{4pt}{
\begin{tabular}{l | lll }
 \bottomrule
\hline
 &{\bf ALiShop-tag}&{\bf ML-25M} &{\bf ML-20M} \\
\hline
\# of users         & 465,573 & 160,775 & 136,677 \\
\# of items         & 1,02,746 & 38,715 & 20,660  \\
\# of interaction   & 4.4M & 12.5M & 10.0M \\
\# of tags          & 263 & 1086 & 1086 \\
\# of tagged items  & 1,02,746 & 29,133 & 13,025 \\
avg. \# of tags per item & 4.16 & 12.61 & 13.46 \\
\# of available tags in $\mathbf{Y}$ & 263 & 1086 & 1086 \\
\toprule
\end{tabular}}
\vspace{-3pt}
\caption{Attributes of datasets after preprocessing.}
\vspace{-13pt}
\label{tab:data_stat}
\end{table}

\subsection{Experimental Settings}

\textbf{Datasets} We experimented with two publicly accessible MovieLens data sets\footnote{The MovieLens data set: \url{https://grouplens.org/datasets/movielens/}} MovieLens-25M and MovieLens-20M, as well as an industrial internal dataset from Alibaba. For both MoviesLens data and Alishop dataset, we regard tags of a movie or an item as its attributes. In this section ``tag'' and ``attribute'' refer to the same thing and will be used interchangeably. For the user movie rating data, we follow MacridVAE, in which ratings are binarized by keeping ratings of four or higher and users who have watched at least five movies. For the tag data, we clean the user provided tags and keep those appeared in the official genome tag table. Additionally we collect a dataset, named AliShop-tag, from Alibaba's e-commerce platform Taobao.  All items in AliShop-tag has tags as well as titles and images. Every user in this dataset clicks at least ten items. The characteristics of the three datasets are summarized in Table\ref{tab:data_stat}. Note, to show all attributes can be considered by the modification data $\mathbf{Y}$ even we restrict $\sum_{t\in\mathcal{T}} |{y}^t_{i,i'}| = 1$, we analyze the number of available tags in $\mathcal{T}$, where a tag $t$ is called available if $\sum_{i,i' \in [1,N]} |{y}^t_{i,i'}| > 1$.\\
\textbf{Query Construction} Queries are created as following: pairs of products that have one attribute difference in their descriptions are selected as the query item and target item pairs; and the modification query is composed by a modification action word (``more'' or ``less'') and the different attribute, e.g. more floral. By this way, triple data is constructed, where the head is a reference item, the tail is a target item and the middle is the modification query. As conventional practice, we hold 20\% of triple data for testing and 80\% for training. The constructed data is close to what will be used in the real-world scenario, where possible modifications will be made offline for each item and then be prompted to a customer who browsed the item just now.

\textbf{Baselines.} We introduce a set of baselines in our experiments:\\
\textit{1. Content-based item Retrieval(CBIR):} We train a fully connected network to predict a matching score between a query (a reference item an a modification) and items. We embed item representation from its interaction history with all users. Besides, we encode the query as a concatenation of the representation of a reference item and the modification text.\\
\textit{2. Text Image Residual Gating \cite{Vo19TIRG} (TIRG):} We adapted this method for item retrieval. TIRG encodes the interaction between one item and users to the item representation. The method aims to map the representation of item and representation of text into the same space and combine them through residual connection. Then we estimate the matching score between the target item and the combination between reference item and modification.\\
\textit{3. Deep Supervised Cross-modal Retrieval \cite{Zhen19DSCMR} (DSCMR):}We adapt this cross-modal text-image retrieval method to our setting. As previously, we use the interaction history of each item as an input feature, and tag texts and text input. This method tries to find a common representation space, in which the samples from different modalities can be compared directly.\\
We also introduce two variants for ablation study to analyze the impact of different components of \CGIR~ to the performance.\\
\textit{1. \CGIR~ w/o VAE:} In this method, instead of using VAE as item representation encoder, we use interaction history as input and a fully connected network as an encoder to encode the interaction history of an item as its representation. This part is same as those baselines. For the remaining, we keep it same as original \CGIR.\\
\textit{2. \CGIR~ w/o Sparse:} We drop the partial sparsity loss and average sparsity loss as shown in \ref{equ:sparseloss}. For the other parts, we keep them same as \CGIR.\\

\begin{table*}[h!]
\centering
\setlength{\tabcolsep}{4pt}{
\begin{tabular}{l | llll | llll | llll }
  \bottomrule
& \multicolumn{4}{c|}{\bf ALiShop-tag} & \multicolumn{4}{c|}{\bf ML-25M}& \multicolumn{4}{c}{\bf ML-20M}\\
\hline
Models &{\bf MGS}&{\bf MGS-C} &{\bf MGS-R}&{\bf Ind.}  &{\bf MGS}&{\bf MGS-C} &{\bf MGS-R}&{\bf Ind.} &{\bf MGS}&{\bf MGS-C} &{\bf MGS-R}&{\bf Ind.} \\
\hline
CBIR   & 0.1811 & 0.2765 & 0.7638 & 0.7627 & 0.1629 & 0.2481 & 0.7904 & 0.6944 & 0.1494 & 0.2575 & 0.7619 & 0.6791 \\
DSCMR  & 0.1934 & 0.2919 & 0.7311 & 0.7398 & 0.1683 & 0.2634 & 0.7819 & 0.6819 & 0.1543 & 0.2763 & 0.7534 & 0.6637 \\
TIRG   & 0.2364 & 0.3566 & 0.7193 & 0.7341 & 0.1903 & 0.2899 & 0.7403 & 0.6563 & 0.2085 & 0.3059 & 0.7264 & 0.6440 \\
\hline
\CGIR & \textbf{0.2796} & \textbf{0.3874} & \textbf{0.8329} & \textbf{0.9834} & \textbf{0.2588} & \textbf{0.3371} & \textbf{0.8961} & \textbf{0.9563} & \textbf{0.2359} & \textbf{0.3516} & \textbf{0.8674} & \textbf{0.9521} \\
\CGIR~ w/o VAE & 0.2314 & 0.3230 & 0.7893 & 0.7692 & 0.2094 & 0.2917 & 0.7388 & 0.6691 & 0.1938 & 0.2972 & 0.7309 & 0.6529 \\
\CGIR~ w/o Sparse & 0.2607 & 0.3841 & 0.8114 & 0.9759 & 0.2361 & 0.3358 & 0.8755 & 0.9312 & 0.2162 & 0.3023 & 0.8448 & 0.9446 \\
\toprule
\end{tabular}}
\vspace{-3pt}
\caption{Gradient Effect. To analysis the gradient effect, we provide a more 
comprehensive analysis using different metrics.} 
\vspace{-13pt}
\label{tab:ConsistRestrict}
\end{table*}

\textbf{Evaluation Metrics.}
We use the following metrics to evaluate the performance of our proposed model. We use two commonly used evaluation criteria in our experiments to evaluate the performance of item retrieval. \textbf{Hit Rate} at K (HR@K) computed as the percentage of test queries where target item is within the top K retrieved items. \textbf{Mean Reciprocal Rank} (MRR) measures the mean of reciprocal rank of target item in the retrieved list. Besides, to qualitatively measure the gradient effect of retrieval result, we design a new metric, named \textbf{Mean Gradient Score} (MGS), to evaluate the degree of gradient for retrieved item list. We use the following equation to define the Mean Gradient Score:
\begin{equation}
\begin{aligned}
MGS & = \frac{1}{|\textit{test}|} \sum_{({i}, \alpha {t}) \in \textit{test}} \Big( Consistency\_Score(Seq\text{-}{{i}}, {t}) \\
 & \cdot \big( 1 - \frac{1}{|\mathcal{T}|} \sum_{\substack{{t}' \in \mathcal{T} \\ {t}' \neq {t}}} Restrictiveness\_Score( Seq\text{-}{i} , {t}') \big)\Big)
\end{aligned}
\label{equ:mgs}
\end{equation}
Here, $test$ is a set of testing pairs. Each testing pair includes an item ${i}$ and an desired modification $\alpha {t}$. $seq\text{-}{i}$ is an item sequence retrieved by increasing the strength coefficient $\gamma$ by 0.1 in each step. The first term in equation \ref{equ:mgs} is the consistency score of the retrieved item sequence. It measures whether the relevant score of items in sequence $Seq\text{-}{i}$ with respect to a certain attribute changes gradually. The second term is the restrictiveness score of the retrieved item sequence. It measures whether the modification on one attribute will influence the relevance between other attributes and items. 
We define the $Consistency\_Score$ and $Restrictiveness\_Score$ as the following:
\begin{equation}
\begin{aligned}
Co&nsistency\_Score(seq\text{-}i,t) = \Big(\frac{1}{N-1}\sum_{k=1}^{K-1} \\ &\vmathbb{1}\big[ \alpha \cdot Relevance(seq\text{-}{i}@k, {t}) < \alpha  \cdot Relevance(seq\text{-}{i}@k+1, {t})\big] \Big)
\end{aligned}
\label{equ:mgs_c}
\end{equation}
\begin{equation}
\begin{aligned}
& Restrictiveness\_Score(seq\text{-}i,t) = 1 - \Big(\frac{1}{N-1}\sum_{k=1}^{K-1} \\ & f\big( \alpha \cdot Relevance(seq\text{-}{i}@k, {t}) < \alpha  \cdot Relevance(seq\text{-}{i}@k+1, {t}) \big)\Big)
\end{aligned}
\label{equ:mgs_r}
\end{equation}
where $N$ is the length of the retrieved sequence $seq\text{-}i$, $seq\text{-}i@k$ is the $k$-th item of the sequence. Specifically, $seq\text{-}i@k$ is the top one item retrieved by the combination of the reference item representation and the modification with scaling coefficient $\gamma = 0.1 \times n$, the retrieved sequence $seq\text{-}i$ is formed by increasing the coefficient. $Relevance(i,t)$ is to calculate the relevance score between item $i$ and tag $t$. For ``add/more'' modification on certain tag, we expect the next item in the gradient sequence to have a higher relevance score regarding tag $g$. For ``remove/less'' modification, we expect a decrease in relevance score. Function $\vmathbb{1}[\cdot]$ is an indicator function which map True and False to 1 and 0. And function $f(\cdot)$ map True and False to 1 and -1. For Restrictiveness Score, if the relevance between indicated tag and items of retrieved sequence in a random walk manner, it will converge to 1 as the length of sequence go to infinite. Note, for MovieLens dataset, a ground-truth relevance score between a movie and a tag is provided. The \textit{Relevance} function directly output the ground-truth relevance score. However, for Alishop-tag dataset, labeling relevance for each item over every attribute is impossible. Therefore, we adopt a heuristic method. For each modification coefficient $\gamma$, instead of measuring the real relevance between top-one retrieved item and a certain attribute, we use the occurrence ratio of items, which has the attribute, on top 100 as the relevance score.

\subsection{Gradient Item Retrieval Performance}
To answer the first and second research questions, we conduct gradient item retrieval on AliShop-tag, MovieLens-20M and MovieLens-25M. The result is shown in table \ref{tab:IR_performance}. 

\textbf{Item Retrieval Performance.} We observe that our approach outperforms the baselines significantly. This is likely because the user interaction is noisy. Directly using interactions as fingerprint for items will include those noise. However, our method use a VAE \cite{ChenB20PairwiseVAE} framework to extract information from user-item interaction. Interpreting from the information bottleneck view \cite{TishbyZ15DLIB}, the disentanglement loss enforces our model to forget those noisy part of data and compress those useful information. Therefore, the noisy user-item interaction is denoised by our method, which gives a high-quality item representation. We also notice that both our method and baselines have a drop on the AliShop-tag dataset. The main reason is likey because the industrial E-commercial dataset is more noisy which will influence the quality of item representation and the item set is larger which directly influences the evaluation metrics because we fix the number $N$ in our experiment. Moreover, we observed that \CGIR~ and TIRG outperfrom CBIR and DSCMR by a significant margin. The improvement is likely because both our method and TIRG use a ranking loss, whereas CBIR and DSCMR use a matching loss.

\textbf{Gradient Retrieval Performance.} To measure the gradient retrieval performance, we apply a modification on a source item representation by increasing/decreasing the strength coefficient $\gamma$ by 0.1 at each time. By analyzing the retrieved item sequences, we calculate the mean gradient score. We outperform all the other baselines methods on MGS. On AliShop-tag data, we achieve better mean gradient score. This is likely because the AliShop-tag dataset has larger number of items which can have a better coverage in the item representation space. During inference stage, less irrelevant items will be retrieved.

\textbf{Ablation Study for Gradient Item Retrieval.} We observe that without using VAE for disentangled item representation, there is a drop on both item retrieval performance and gradient retrieval performance and the impact on gradient retrieval performance is more serious. One reason is attribute-relevant information will appear at each dimension of distributional item representations. When a scaled modification is applied, more than one attributes' information will be changed. Another reason as we discuss previously, the VAE structure delivers a denoised item representation. Besides, we observe that without using sparse loss the item retrieval performance is competitive with the best \CGIR~, but the gradient item retrieval performance is affected obviously. This result is in the line with our expectation. The sparse loss will compress the semantic meaning of an attribute representation to several dimensions which avoid a modification on irrelevant attributes, however some information will be lost in the meanwhile.

\subsection{Gradient Effect Study}
In order to analyze the effect of disentanglement and answer the third research question. We provide two more detailed experiments. In the first one, we measure the consistency and restrictiveness. In the second one, we analyze the relation between independence level and mean gradient score(MGS).

\textbf{Consistency and restrictiveness}
To validate the effect of disentangled representation, we measure the consistency and restrictiveness separately. More specifically, we calculate the mean value of restrictiveness score (equation \ref{equ:mgs_r}) and consistency score (equation \ref{equ:mgs_c}). Additionally, We quantify the level of independence by calculating the Uncorrelatedness of item representations \cite{mazhou0Y019MacridVAE}. We define Uncorrelatedness as: 
\begin{equation}
\begin{split}
Ind\_level(\mathbf{H}) = 1 - \frac{1}{D(D-1)}\sum_{\substack{d_i, d_j \in [1,D] \\ d_i \neq d_j}  } \big|CorrCoef ( {\mathbf{H}_{:,d_i}, \mathbf{H}_{:,d_j}}\big ) |. 
\end{split}
\end{equation}
The function $CorrCoef()$ measures the correlation coefficient between two variables. As shown in table \ref{tab:ConsistRestrict}, we denote restrictiveness score as \textbf{MGS-R}, mean consistency score as \textbf{MGS-C} and independence level as \textbf{Ind.}. 

We observed that the independence level outperform all other methods, which indicates that our method can achieve factorized item representation. This  directly answer the research question 3. We also observed that \CGIR~ outperform other competitor on \textbf{MGS-R}. This improvement shows that \CGIR~ has less influence on irrelevant hidden factor, when one factor was changed. This main reason is likely because by applying the disentangled loss, the item representation is factorized, so different hidden factors are encoded into different dimensions of the item representation, which allows us to only modify the value a few dimensions during inference. The performance on metric independence level also supports this explanation. Besides, we notice that although both VAE structure and sparse loss can impact the consistency and restrictiveness, the VAE is more important for important for disentangled representation.

\begin{figure}[ht]
\centering
\includegraphics[width=0.45\textwidth]{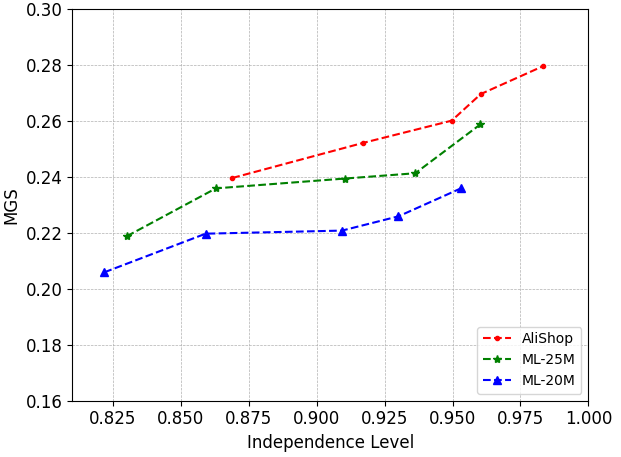}
\caption{\textbf{Independence Level vs. Mean Gradient Score}}
\label{fig:ind-mgs}
\end{figure}

\textbf{Independence Level and Mean Gradient Score}
In order to analyze the relation between Mean Gradient Score and Independence Level achieved by our disentangled representation. We vary the hyper-parameters related with disentanglement ($\beta$ and $\rho$ for our method), and plot Figure \ref{fig:ind-mgs} the relationship between the level of independence and Mean Gradient Score. We use all item representations on all three datasets to calculate the level of independence. By improving independence of item representations, we achieve a better result on gradient retrieval. This suggests that disentanglement loss can help improve the gradient item retrieval result.

\begin{figure}[ht]
\centering
\begin{subfigure}[b]{0.45\textwidth}
   \includegraphics[width=1\linewidth]{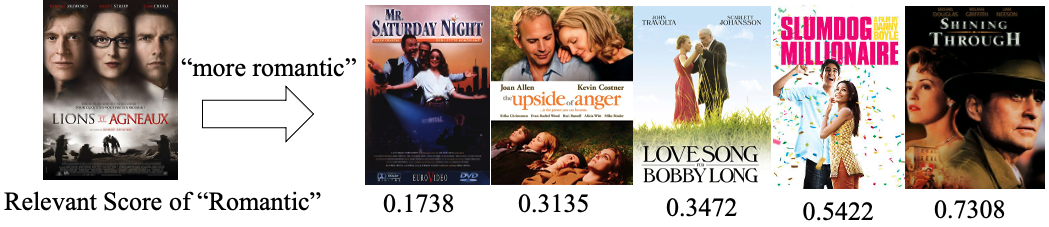}
   \caption{Apply modification "more romantic" on movie Lions For Lambs (2007)}
   \label{fig:ml_increase} 
\end{subfigure}

\begin{subfigure}[b]{0.45\textwidth}
   \includegraphics[width=1\linewidth]{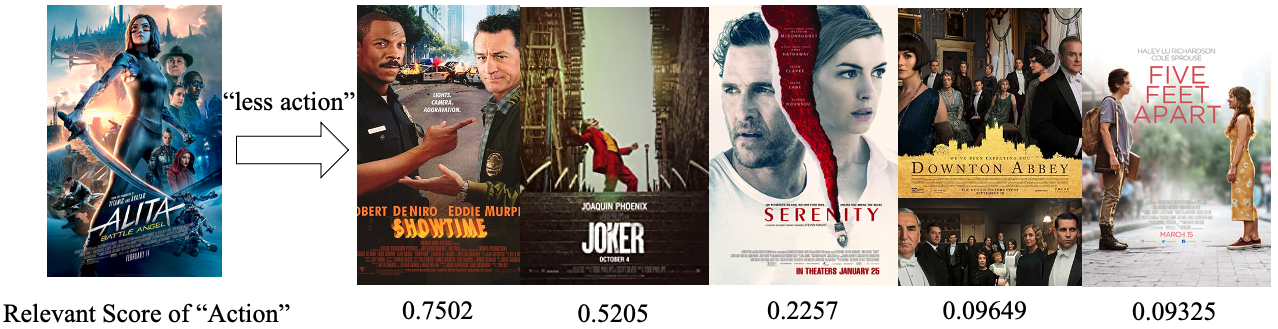}
   \caption{Apply modification "less action" on movie Alita:Battle Angel(2019)}
   \label{fig:ml_decrease}
\end{subfigure}
\caption{Apply modification on movie data to retrieve a sequence of movies in gradient manner.}
\label{fig:ml_visual}
\end{figure}

\begin{figure}[ht]
\centering
\begin{subfigure}[b]{0.45\textwidth}
   \includegraphics[width=1\linewidth]{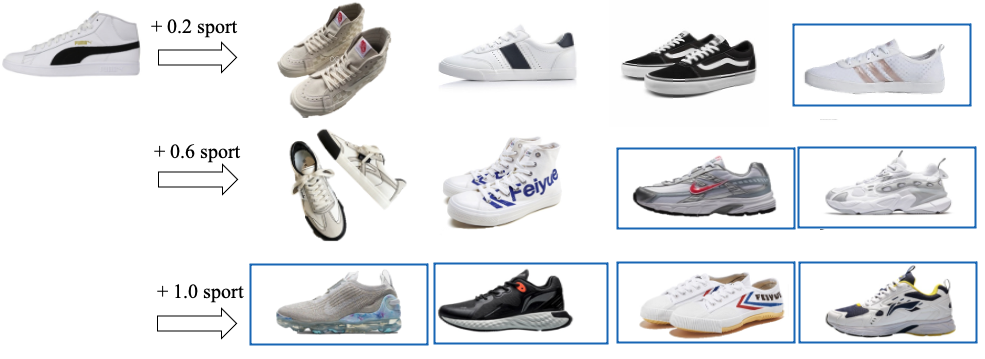}
   \caption{Apply modification "more sport" on a sneaker}
   \label{fig:ali_sport}
\end{subfigure}

\begin{subfigure}[b]{0.45\textwidth}
   \includegraphics[width=1\linewidth]{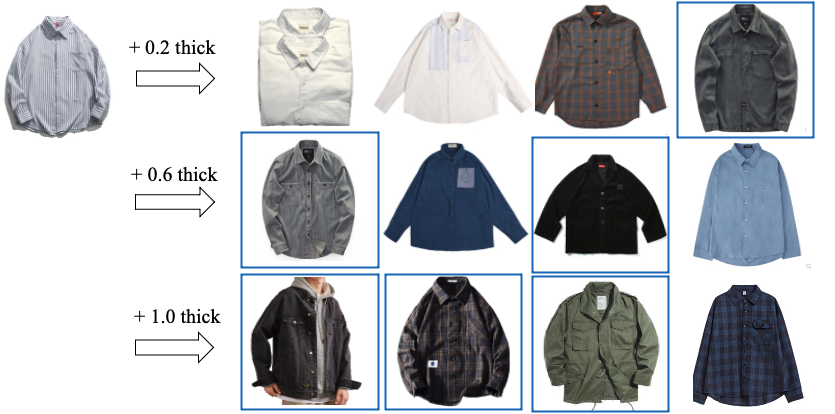}
   \caption{Apply modification "more thick" on a light shirt}
   \label{fig:ali_thick}
\end{subfigure}
\caption{Apply modification on Alishop-tag data, a sequence of items are retrieved after changing the value of $\gamma$ each time}
\label{fig:ali_visual}
\end{figure}

\textbf{Case Study}
To illustrate the gradient effect achieved by our method, we visualize several cases as shown in Figure \ref{fig:ml_visual} and \ref{fig:ali_visual}. For MovieLens datasets, because the ground truth relevance score between moives and movie tags are given, we show the relevance score under the poster of each movie. We retrieved those movies by changing the value of modification strength coefficient $\gamma$ from 0.2 to 1.0 increasing 0.2 at each step. We only keep the top 1 movie into the gradient item retrieval list for each $\gamma$. We visualize ``more'' and ``less'' modification results in Figure \ref{fig:ml_increase} and Figure \ref{fig:ml_decrease}, respectively. For Alishop-tag dataset, we show the top 4 items retrieved by different modification strength coefficients. This is because we do not have ground-truth relevance score between tags and products. A heuristic way to measure and show the relevance is to use the number of desired products appeared in top@K as the relevance score. As shown in Figures \ref{fig:ali_sport} and \ref{fig:ali_thick}, the number of desired items in top 4 of each retrieved list is increasing.

%% file: sec-related.tex
\section{related work}
\label{sec:Related Work}
\textbf{Product Search and Item Retrieval}
There are a lot of research has been done on product searches by incorporating text into the query, such as \cite{JiangHX12MMdetection, HanWHZZLZD17ConceptDiscovery} use the user's feedback to the search query. For the problem of image-based product search Vo et al. \cite{Vo19TIRG} proposed a method that regards an image and a text string as a query and allows attribute modification. Besides, for image-based fashion search, Zhao et al. \cite{Zhao17MemoryFashionSearch} developed a memory-augmented deep learning system that can perform
attribute manipulation based on the reference image. Moreover, there are a lot of cross-modal methods that deal with the item retrieval problem \cite{Zhen19DSCMR, wen19advCrossModalIR, Yu19UnsupCrossModalIR}. Cross-modal methods try to encode information from different modalities into a common representation space. To learn a high-quality common representation space, Zhen et al. \cite{Zhen19DSCMR} leveraged data pairs to match them in representation space. To deal with the unparalleled data scenario, \cite{wen19advCrossModalIR} proposed an adversarial learning method to deal with it. We are approaching the item retrieval problem where image data are not available. Besides, unlike previous work which seldom shows its effectiveness of gradient retrieval, in this work, we also focus on how to retrieval items in a gradient manner with respect to certain attributes indicated by a modification.
\\
\textbf{Disentangled representation learning} 
Disentanglement is an open problem in the realm of representation learning which aims to identify and
disentangle the underlying explanatory factors \cite{Bengio13Disentangle}. There are a lot of works that focus on unsupervised disentanglement \cite{Higgins17betaVAE, kumar18DIPVAE, Zhao17InfoVAE, chen18betaTCVAE, KimM18FactorVAE, mazhou0Y019MacridVAE}. $\beta$-VAE \cite{Higgins17betaVAE} demonstrates that disentanglement can emerge once the KL divergence term in the VAE \cite{KingmaW13VAE} objective is aggressively penalized. Later, Zhao et al. \cite{Zhao17InfoVAE} proposed InfoVAE which regarded VAE from the view of information theory. By maximizing the mutual information between the data variables and latent variables, the mutual information between the latent variables is minimized. However, Locatello et al. \cite{Locatello18ChallengeDisentangle} theoretically demonstrate that unsupervised learning of disentanglement arises from model inductive bias and empirically shows that many existing methods for the unsupervised learning of disentangled representations are brittle, requiring careful supervision-based hyper-parameter tuning. Therefore, recently, the research attention has turned to forms of disentanglement in supervised or weakly supervised setting \cite{Feng18DualSwap, Chen17MVGAN, ChenB20PairwiseVAE, Gabbay19LatentOptimDisentangle}. To model pairwise similarities between data samples, Chen et al.\cite{ChenB20PairwiseVAE} proposed a pairwise VAE that tries to capture a binary relationship(similar or not). And Feng et al.\cite{Feng18DualSwap} proposed a Dual Swap Disentangling method to leverage binary similarity labels. Besides, a theoretical framework was given by Shu et al. \cite{shu20disentangleguarantee}, which guarantees consistency and restrictiveness can be achieved under three types of weakly supervised setting. Different from \cite{Feng18DualSwap, ChenB20PairwiseVAE}, in this work, we focus on using ranking triples information as supervision. In the ranking triples, not only the ranking relation between two data samples were given, but we also provided the information about we compare the two data samples in which point of view. Besides, our method aims to ground the semantic meaning of the comparison view into the dimensions of disentangled representations.
\\
\textbf{Critiquing Recommender Systems}
Critiquing is a method widely used conversational recommendation\cite{thompson2004personalized} which supports a task-oriented, multi-turn dialogue with their users to discover the detailed and current preferences of the user\cite{jannach2020survey}. In critiquing approaches, users are presented with a recommendation result during the dialogue and then apply pre-defined critiques on the result\cite{burke1997findme, hammond1995faq}. Specifically, in this setting, a user is iteratively provided with an item
recommendation and attribute description for that item; a user
may either accept the recommendation, or critique the attributes
in the item description to generate a new recommendation result\cite{tou1982rabbit}. Recently, there are some works introduce the critiquing method into the current deep learning recommendation system to improve the explainability of the system\cite{antognini2020interacting} which a system proposes to the user a recommendation with its keyphrases and the user can interact with the explanation and critique phrases. Furthermore, there are some work focus on the latent linear critiquing\cite{luo2020latent, luo2020deep} which built on  existing linear embedding recommendation
algorithm to co-embed keyphrase attributes and user preference embeddings and modulate the strength of multi-step critiquing feedback. By levering the linear structure of the embeddings, the number of interactions required to find a satisfactory item is reduced. Different from those methods, we think a better way is to provide a user with an item sequence with a gradual change on an indicated attributes in order to allow users to obtain satisfactory items with as few interactions as possible. Besides, in those methods, keyphrase frequency usage data is necessary to learn the strengthen of a critiquing. However, in our method, only attributes data is required.

%% file: sec-conclusion.tex
\section{conclusion}
\label{sec:conclusion}
In this paper, we identify and study a new problem -- gradient item retrieval. It is defined as retrieving a sequence of items with gradual change with respect to a certain attribute indicated by a modification text. To solve this problem, we proposed a novel method Controllable Gradient Item Retrieval \CGIR. Our method takes a product and a modification text, which indicates what attributes to change and how to change, as a query and retrieves a sequence of items with gradual change on the relevance between the indicated tag and items in the sequence. To achieve the gradient effect, our method learns a disentangled item representation with weak supervision and grounds semantic meanings to dimensions of the representation. We show that our method can achieve consistency and restrictiveness under a previously proposed theoretical framework. Empirically, we demonstrate that our method can retrieve items in a gradient manner; and in item retrieval tasks, our method outperforms existing approaches on three different datasets.

%% file: sec-acknowledge.tex
\section{acknowledgement}
\label{sec:acknowledge}

This work is supported by National Science Foundation under Award No. IIS-1947203 and IIS-2002540. The views and conclusions are those of the authors and should not be interpreted as representing the official policies of the funding agencies or the government.